# Multiscale modeling meets machine learning:
# What can we learn?

Grace C.Y. Peng[a], Mark Alber[b], Adrian Buganza Tepole[c], William Cannon[d], Suvranu De[e], Salvador Dura-Bernal[f], Krishna Garikipati[g], George Karniadakis[h], William W. Lytton[f], Paris Perdikaris[i], Linda Petzold[j], Ellen Kuhl[k,*]

[a]*National Institute of Biomedical Imaging and Bioengineering, National Institutes of Health, Bethesda, Maryland, USA*
[b]*Department of Mathematics, University of California, Riverside, USA*
[c]*Department of Mechanical Engineering, Purdue University, Lafayette, Indiana, USA*
[d]*Computational Biology, Pacific Northwest National Laboratory, Richland, Washington, USA*
[e]*Department of Mechanical, Aerospace and Nuclear Engineering, Troy, New York, USA*
[f]*Department of Physiology and Pharmacology, State University of New York, New York, USA*
[g]*Departments of Mechanical Engineering, and Mathematics, MICDE, University of Michigan Ann Arbor, Michigan, USA*
[h]*Division of Applied Mathematics, Brown University, Providence, Rhode Island, USA*
[i]*Department of Mechanical Engineering, University of Pennsylvania, Philadelphia, Pennsylvania, USA*
[j]*Department of Computer Science and Mechanical Engineering, University of California, Santa Barbara, California, USA*
[k]*Department of Mechanical Engineering, Stanford University, Stanford, California, USA*

**Abstract**

Machine learning is increasingly recognized as a promising technology in the biological, biomedical, and behavioral sciences. There can be no argument that this technique is incredibly successful in image recognition with immediate applications in diagnostics including electrophysiology, radiology, or pathology, where we have access to massive amounts of annotated data. However, machine learning often performs poorly in prognosis, especially when dealing with sparse data. This is a field where classical physics-based simulation seems to remain irreplaceable. In this review, we identify areas in the biomedical sciences where machine learning and multiscale modeling can mutually benefit from one another: Machine learning can integrate physics-based knowledge in the form of governing equations, boundary conditions, or constraints to manage ill-posed problems and robustly handle sparse and noisy data; multiscale modeling can integrate machine learning to create surrogate models, identify system dynamics and parameters, analyze sensitivities, and quantify uncertainty to bridge the scales and understand the emergence of function. With a view towards applications in the life sciences, we discuss the state of the art of combining machine learning and multiscale modeling, identify applications and opportunities, raise open questions, and address potential challenges and limitations. This review serves as introduction to a special issue on Uncertainty Quantification, Machine Learning, and Data-Driven Modeling of Biological Systems that will help identify current roadblocks and areas where computational mechanics, as a discipline, can play a significant role. We anticipate that it will stimulate discussion within the community of computational mechanics and reach out to other disciplines including mathematics, statistics, computer science, artificial intelligence, biomedicine, systems biology, and precision medicine to join forces towards creating robust and efficient models for biological systems.

# 1. Motivation

Machine learning is rapidly infiltrating the biological, biomedical, and behavioral sciences and seems to hold limitless potential to transform human health [127]. It already is widely considered to be one of the most significant breakthroughs in medical history [28]. But can this technology really live up to its promise? Machine learning is the scientific discipline that seeks to understand and improve how computers learn from data. As such, it combines elements from statistics, understanding relationships from data, with elements from computer science, developing algorithms to manage data. The success of machine learning relies heavily on our ability to collect and interpret big data. In many fields of medicine, we have successfully done this for multiple decades. So what's really new? The recent excitement around machine learning is generally attributed to the increase in computational resources, cloud storage, and data sharing, which we can witness in our own lives through smart watches, wearable electronics, or mobile devices [45]. For example, a recent success story of machine learning in medicine has shown that it is possible to classify skin cancer into malignant and benign subtypes using photographic images, for example from smartphones [34]. Unarguably, the two most compelling opportunities for machine learning in biomedicine are diagnosis and prognosis. Potential applications range from identifying bone fracture, brain hemorrhages, and head trauma to detecting lung nodules, liver masses, and pancreatic cancer [127]. But is machine learning powerful and accurate enough that we can simply ignore physics based simulations entirely?

Machine learning is exceptionally good at integrating multimodality multi-fidelity data with the goal to reveal correlations between different features. This makes the technology very powerful in fields like radiology and pathology where we seek to classify risk or stratify patients based on medical images [127] and the answer is either binary, as in the classification of skin cancer [34], or a discrete number, as in a recent classification of 12 types of arrhythmia [43]. In fact, arrhythmia classification is an excellent example, because more than 300 million electrocardiograms are acquired annually worldwide, and we have access to a vast amount of annotated data. Problems may arise, however, when dealing with sparse or biased data [102]. In these cases the naive use of machine learning can result in ill-posed problems and generate non-physical predictions. Naturally, this brings up the question that, provided we know the underlying physics, can we integrate our prior knowledge to constrain the the space of admissible solutions to a manageable size [100]?

Recent trends in computational physics suggest that exactly this approach [14]: to create data-efficient physics-informed learning machines [96, 97]. Biomedicine has seen the first successful application of these techniques in cardiovascular flows modeling [50] or in cardiac activation mapping [109], where we already have a reasonable physical understanding of the system and can constrain the design space using the known underlying wave propagation dynamics. Another example where machine learning can immediately benefit from multiscale modeling and physics-based simulation is the generation of synthetic data [106], for example, to supplement sparse training sets. This raises the obvious question–especially within the computational mechanics community– where can physics-based simulations benefit from machine learning?

Physics-based simulations are enormously successful at integrating multiscale, multiphysics data with the goal of uncovering mechanisms that explain the emergence of function [19]. In biomedicine, physics-based simulation and multiscale modeling have emerged as a promising technologies to build organ models by systematically integrating knowledge from the molecular, cellular, and tissue levels [26] as evidenced by initiatives like the United States Federal Interagency Modeling and Analysis Group IMAG [138]. Two immediate opportunities for machine learning in multiscale modeling include learning the underlying physics [104] and learning the parameters for a known physics-based problem. Recent examples of learning the underlying physics are the data-driven solution of problems in elasticity [23] and the data-driven discovery of partial differential equations for nonlinear dynamical systems [14, 95, 98]. This class of problems holds great promise, especially in combination with deep learning, but involves a thorough understanding and direct interaction with the underlying learning machines [100]. Are there also immediate opportunities for integrating machine learning and multiscale modeling, more from an end-user perspective, without having to modify the underlying tools and technologies at their very core?

This article seeks to answer the question of how multiscale models can benefit from machine learning [3]. It is the introduction to a special issue on 'Uncertainty Quantification, Machine Learning, and Data-Driven Modeling of Biological Systems' that seeks to place this theme within the broader field of computational mechanics. The most common example, in this special issue alone, is the theme of learning model parameters, ranging from characterizing brain stiffness [124] to cardiac conductivity [9]. Another immediate application in biomedicine is using machine learning to create fast and robust surrogate models, as discussed in this special issue for the examples of cardiac electrophysiology [107], vascular hemodynamics [39], and skin growth [61]. As we will see, these surrogate models allow us to seamlessly integrate multimodality or multi-fidelity data. We can also use neural networks to directly map pressure and topology as input onto the deformation field as output, for example to accelerate decision making in electrosurgery [42]. Finally, natural questions that machine learning can help us answer focus on sensitivity analysis [62] and uncertainty quantification [61, 146]. We have structured this introduction to the special issue around four methodological areas, ordinary and partial differential equations, and data and theory driven machine learning [3]. For each area, we discuss the state of the art, identify applications and opportunities, raise open questions, and address potential challenges and limitations in view of specific examples from the life sciences. To make this work accessible to a broad audience, we summarize the most important terms and technologies associated with machine learning in boxes where they are first mentioned. We envision that this work will stimulate discussion and inspire scientists in the broader field of computational mechanics to explore the potential of machine learning towards creating reliable and robust predictive tools for biological, biomedical, and behavioral systems to the benefit of human health.

## 2. Ordinary differential equations

Ordinary differential equations in time are ubiquitous in the biological, biomedical, and behavior sciences. At the molecular, cellular, organ, or population scales it is often easier to make observations and acquire data associated with ordinary differential equations than for partial differential equations, since the latter encode spatial variations, which are often more difficult to access. Ordinary differential equation based models can range from single equations to large systems of equations or stochastic ordinary differential equations. This implies that the number of parameters is typically large and can easily reach hundreds or more. Figure 1 illustrates an example of ordinary differential equations to explore the biophysical mechanisms of development [126].

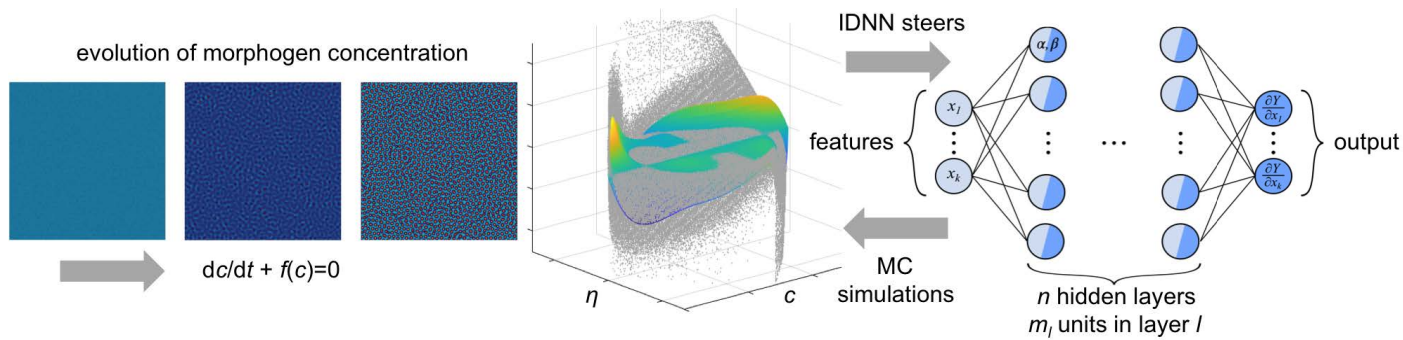

Figure 1: **Ordinary differential equations.** Biophysical mechanisms of development can be discerned by identifying the nonlinear driving terms in ordinary differential equations that govern the evolution of morphogen concentrations, left. Metabolic processes evolve on a free energy landscape $g(c, \eta)$ that can be explored by Monte Carlo simulations, thus generating large scale data, shown as a grey point cloud, that are used to train various classes of neural networks, right. The colored surface is an integrable deep neural network [126] representation of the metabolic data.

## 2.1. State of the art

Assuming we have acquired adequate data, the challenge begins with identifying the nonlinear, coupled driving terms. To analyze the data, we can apply formal methods of *system identification*. Common techniques include classical regression using L1 or LASSO and L2 or ridge regression, as well as stepwise regression with statistical tests [14, 134]. These approaches are essentially nonlinear optimization problems that learn the set of coefficients by multiplying combinations of algebraic and rate terms that result in the best fit to the observations. For adequate data, the *system identification* problem is usually relatively robust and can learn a parsimonious set of coefficients, especially with stepwise *regression*. Clearly, parsimony is central to identifying the correct set of equations and the easiest strategy to satisfy this requirement is classical or stepwise *regression*.

> ***System identification*** *refers to a collection of statistical methods that identify the governing equations of a system from data. These methods can be applied to obtain either an equilibrium response or the dynamics of a system. Typical examples include inferring operators that form ordinary [73] or partial [134] differential equations.*

> ***Regression*** *is a supervised learning approach in which the algorithm learns from a training set of correctly identified observations and then uses this learning to evaluate new observations. The underlying assumption is that the output variable is continuous over the input space. Examples in biomedicine include predicting an individual's life expectancy, identifying a tolerable dose of chemotherapy, or exploring the interplay between drug concentration and arrhythmogenic risk [106], among many others.*

Any discussion of *system identification* from experimental data should address *uncertainty quantification* to account for both measurement errors and model errors. The Bayesian setting provides a formal framework for this purpose [48]. Prior probability distribution functions must be assumed for the errors. In the absence of deeper insights into the measurement techniques, a common choice is the Gaussian distribution. On this note, we observe that recent *system identification* techniques [14, 73, 104, 92, 74, 20, 134] start from a large space of candidate terms in the ordinary differential equations to systematically control and treat model errors. Machine learning provides a powerful approach to reduce the number of dynamical variables and parameters while maintaining the biological relevance of the model [14, 116].

> ***Uncertainty quantification*** *is the science of determining the likelihood of the outputs if the inputs are not exactly known. Since standard variations in biomedical data are usually large and it is critical to know how small variations in the input data affect the output, uncertainty quantification is indispensable in medicine. The inherent nonlinearity of biological process also drives a need for uncertainty analysis, since noise in the inputs is nonlinearly propagated through the system. Sources of uncertainty can be experimental or computational related to the underlying governing equations, their parameters, and boundary conditions. Some examples include quantifying the effects of experimental uncertainties in heart failure [86], the effects of biomechanical stimuli in coronary artery bypass grafts [130], or the effects of material properties on stress profiles in reconstructive surgery [59].*

## 2.2. Applications and opportunities

There are numerous applications of ordinary differential equations that integrate machine learning and multiscale modeling for biological, biomedical, and behavioral systems.

***Metabolic networks.*** Machine learning has been applied to take advantage of large amounts of genomics and metabolomics data for the optimization of ordinary differential equation-based metabolic network models and their analysis [25]. For example, machine learning and genome-scale models were applied to determine the side effects of drugs [114]. Also, a recent study used a combination of machine learning and multiomics data, proteomics and metabolomics, to effectively predict pathway dynamics providing qualitative and quantitative predictions for guiding synthetic biology efforts [24]. *Supervised learning* methods are often used for finding the metabolic dynamics represented by coupled nonlinear ordinary differential equations to obtain the best fit with the provided time-series data.

> ***Supervised learning*** *defines the task of learning a function based on previous experience in the form of known input-output pairs or function evaluations. In many cases, this is a task that a trained person can do well, and the computer is trying to approximate human performance. When the input is high-dimensional, or the function is highly nonlinear, personal intuition may not be useful and supervised learning can overcome this limitation. Typical examples include classification and regression tasks. In biomedicine, a common classification problem is pattern recognition in an electrocardiogram to select from a limited set of diagnoses [43]; other examples include detecting cancer from medical images [34], estimating risk scores for coronary heart disease, guiding antithrombotic therapy in atrial fibrillation, and automating implantable defibrillators in hypertrophic cardiomyopathy [28].*

> ***Unsupervised learning*** *defines the task of identifying naturally occurring patterns or groupings within datasets that consisting of input features without labeled output responses. The most common types of unsupervised learning techniques include clustering and density estimation used for exploratory data analysis to identify hidden patterns or groupings. In biomedicine, a promising example is precision medicine [28].*

***Microbiology, immunology, and cancer.*** The coupled, nonlinear dynamics of intracellular and extracellular signaling are represented by cascades of tens of ordinary differential equations to model the onset of tuberculosis [101]. The same approach is applied to modeling the interaction of the immune system and drugs in the mathematical biology of cancer [79]. In this case, the major challenge is *system identification*. Another application is bridging scales in cancer progression in specific micro-environments by mapping genotype to phenotype using neural networks [37].

***Neuroscience.*** Machine learning is applied to *system identification* of the ordinary differential equations that govern the neural dynamics of circadian rhythms [12, 29, 78]. Principal component analysis and neural networks have been more widely applied to memory formation [84, 91], chaotic dynamics of epileptic seizures [1, 2], Alzheimers Disease, and aging.

***Biomechanics.*** The most prominent potential application of machine learning in biomechanics is in the determination of response functions including stress-strain relations or cell-scale laws in continuum theories of growth and remodeling [4]. These relations take the form of both ordinary differential equations, for which *system identification* is of relevance, and direct response functions, for which the framework of *deep neural networks* is applicable [105]. For example, a recent study integrated machine learning and multiscale modeling to characterize the dynamic growth and remodeling during heart failure across the scales, from the molecular via the cellular to the cardiac level [86].

> ***Deep neural networks*** *are a powerful form of machine learning strategies to approximate functions. The input features proceed through multiple hidden layers of connected neurons that progressively compose these features and ultimately produce an output. The key feature of deep learning is that the architecture of the network is not determined by humans, but rather by the data themselves. Deep neural networks have been successfully used in image and speech recognition [58]. The number of examples of deep learning in biomedicine is rapidly increasing and includes interpreting medical images to classify tuberculosis, identify bone fracture, detect lung nodules, liver masses, and pancreatic cancer, identify brain hemorrhages and head trauma, and analyze mammograms and electrocardiograms [127].*

***Public health.*** The dynamics of disease spreading through a population, affected by environmental factors, has long been represented by cascades of ordinary differential equations. A major challenge in this application is determining the parameters of the ordinary differential equations by *system identification* [18]. Interestingly, the ordinary differential equations of disease spreading have recently been adopted to model the prion-like spreading of neurodegenerative diseases [137], where the parameters could potentially be identified from magnetic resonance images using machine learning.

## 2.3. Open questions

***Maximizing information gain.*** An open question in modeling biological, biomedical, and behavior systems is how to best analyze and utilize sparse data. In such a setting, sparse identification techniques must be integrated with the experimental program. Optimal experimental design [46] methods allow the most efficient choice of experiments to maximize the information gain using criteria such as the Kullback-Leibler divergence or the Akaike Information Criterion. The information theoretic approach is particularly powerful to treat model form errors. In biological systems, where data may be obtained by resource-intensive wet lab experiments or multi-scale model simulations, the most efficient combination of such approaches could result in maximizing novel biological insight.

***Optimizing efficiency.*** The identification of the system of governing equations still leaves open the question of efficiency and time to solution, especially if equations are to be used in a sampling approach such as Monte Carlo. In this setting, the repeated generation of solutions inevitably suggests that we should circumvent the expense of time integration methods. Deep learning methods centered on neural networks offer a number of options. Particularly well-suited are *recurrent neural networks*, based on long short-term memory cells, which account for the inherent time dependence of ordinary differential equations and their solutions. Neural networks can efficiently encode the complex time dependence of ordinary differential equations. For example, extremely high-order time integration schemes such as Runge-Kutta algorithms of the order of one hundred have been replaced successfully by tailored *deep neural networks* [96]. The construction and use of such surrogate models is indispensable for sampling upward of tens of thousands of entire trajectories of dynamical systems such as reaction-diffusion of coupled ligand-morphogen pairs.

> ***Recurrent neural networks*** *are a class of neural networks that incorporate a notion of time by accounting not only for current data, but also for history with tunable extents of memory. A recent application is identifying unknown constitutive relations in ordinary differential equation systems [40]*

***Combining multi-fidelity neural networks.*** Neural networks can be directly trained against labels for a quantity of interest such as the time-averaged solution or its frequency distribution. Principal component analysis can be applied to the dynamics to develop reduced order models. *Deep neural networks* can be combined into an approach of *multi-fidelity learning* [57] that integrates well with multiscale modeling methods. In this approach, multiple neural networks are trained. Coarse scale, but plentiful data, for example obtained from larger numbers of trajectories reported at fewer time instants, are used to train low-fidelity neural networks, which are typically shallow and narrow. Progressively finer scale data at increasing numbers of time instants, but for fewer trajectories and expensive to obtain, are used to train higher fidelity *deep neural networks* to minimize the error between the labels and the output of the low-fidelity neural network. The low-fidelity neural network resolves the low frequency components of the response, with the progressively higher fidelity *deep neural networks* representing the higher frequencies. The underlying principle is that the knowledge base of the response is resolved by shallower and narrower neural networks, while the critical, high frequency response is left to the high-fidelity deep neural network. Developing these novel approaches is important to accurately resolve the dynamics of, for example, reaction-diffusion systems of ligands and morphogens that together control patterning in developmental biology.

*2.4. Potential challenges and limitations*

***Dealing with inadequate resolution.*** The most prominent challenge facing the application of machine learning and data-driven methods to obtaining a better understanding of biological systems is a paucity of data. In ordinary differential equation modeling, we often have to rely on classical data acquisition techniques, for example, microscopy or spectroscopy, which are known to have limited temporal resolution. Obtaining time series data at a sufficiently high temporal resolution to build and train mathematical models has always been and will likely remain a major challenge in the field.

***Processing sparse data.*** Sparse, incomplete, or heterogeneous data pose a natural challenge to modeling biological, biomedical, and behavioral systems. In principle, direct numerical simulations can fill this gap and generate missing data. However, the simulations themselves can be limited by poorly calibrated parameter values. There is, therefore, a pressing need to develop robust inverse methods that are capable of handling sparse data. One example is robust *system identification*, the creation of mathematical models of dynamical systems from measured experimental data. This naturally implies the optimal design of experiments to efficiently generate informative training data and iterative model refinement or progressive model reduction.

## 3. Partial differential equations

Partial differential equations describe the physics that govern the evolution of biological systems in time and space. The interaction between the different scales, both spatial and temporal, coupled with the various physical and biological processes in these systems, is complex with many unknown parameters. As a consequence, modeling biological systems in a multi-dimensional parametric space poses challenges of *uncertainty quantification*. Moreover, modeling these systems depends crucially on the available data, and new multi-modality data fusion methods will play a key role in the effective use of partial differential equation modeling of multiscale biological systems. An additional challenge of modeling complex biological phenomena stems from the lack of knowledge of some of the processes that need to be modeled. Physics-informed machine learning is beginning to play a central role at this front, leveraging *multi-fidelity* or multi-modality data with any known physics. These data can then be exploited to discover the missing physics or unknown processes.

## 3.1. State of the art

Modeling biological, biomedical, and behavioral systems crucially depends on both the amount of available data and the complexity of the system itself. The classical paradigm for which many numerical methods have been developed over the last fifty years is shown in the top of Figure 2, where we assume that the only data available are the boundary and initial conditions, while the specific governing partial differential equations and associated parameters are precisely known. On the other extreme, in the bottom of Figure 2, we may have a lot of data, for example in the form of time series, but we do not know the governing physical law, for example the underlying partial differential equation at the continuum level.

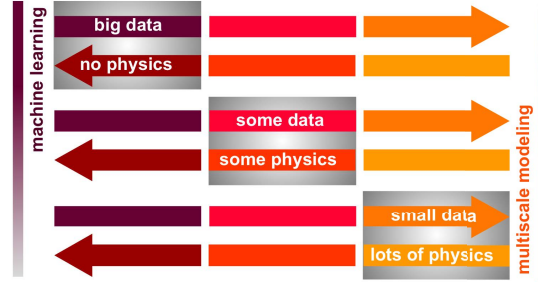

Figure 2: **Categories of modeling biomedical systems and associated available data and underlying physics.** We use the term physics to imply the known physics for the target problem. Physics-informed neural networks can seamlessly integrate data and mathematical models, including models with missing biophysics, in a unified and compact way using automatic differentiation and partial differential equation-induced neural networks.

Many problems in social dynamics fall under this category, although work so far has focused on recovering known partial differential equations from data only. Perhaps the most interesting category for biological systems is sketched in the middle of Figure 2, where we assume that we know the physics partially but not entirely. For example, we know the conservation law and not the constitutive relationship, but we have several scattered measurements in addition to the boundary and initial conditions that we can use to infer the missing functional terms and other parameters in the partial differential equation and simultaneously recover the solution. This middle category is the most general case. In fact, it is representative of the other two categories, if the measurements are too few or too many. This mixed case may lead to the significantly more complex scenarios, where the solution is a stochastic process due to stochastic excitation or an uncertain material property, for example the diffusivity in a tissue. Hence, we can employ stochastic partial differential equations to represent these stochastic solutions and other stochastic fields. Finally, there are many problems involving long-range spatio-temporal interactions, for example the viscoelasticity of arteries or the super-diffusion inside a cell, where fractional calculus and fractional partial differential equations, rather than the currently common partial differential equations with integer order derivatives, may be the proper mathematical model to adequately describe such phenomena.

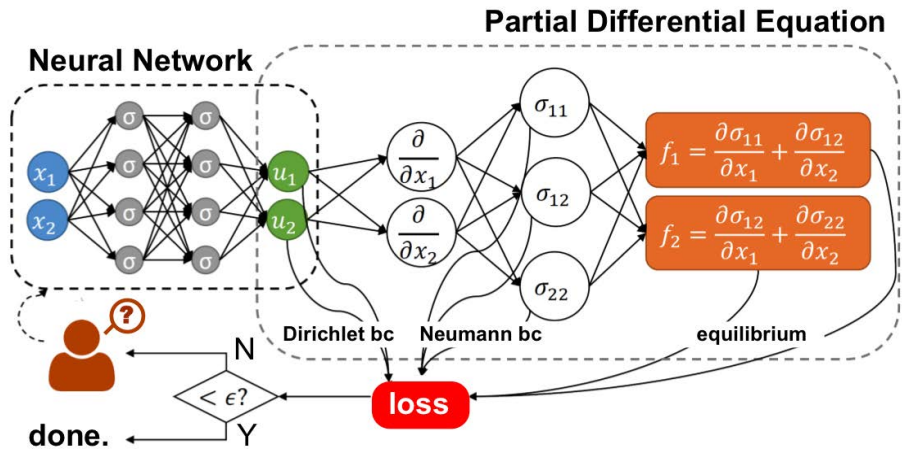

Figure 3: **Physics-informed neural networks.** The left physics uninformed network represents the solution $u(x, t)$ of the partial differential equation; the right physics informed network describes the residual $f(x, t)$ of the partial differential equation. The example illustrates the nonlinear Schrödinger equation with unknown parameters $\lambda_1$ and $\lambda_2$ to be learned. In addition to unknown parameters, we can learn missing functional terms in the partial differential equation. Currently, this optimization is done empirically based on trial and error by a human-in-the-loop. Here, the $u$-architecture is a fully-connected neural network, while the $f$-architecture is dictated by the partial differential equation and is, in general, not possible to visualize explicitly. Its depth is proportional to the highest derivative in the partial differential equation times the depth of the uninformed $u$ neural network.

***Physics-informed machine learning.*** Prior physics-based information in the form of partial differential equations, boundary conditions and constraints regularize a machine learning approach in such a way that it can then learn from small and noisy data that can evolve in time. Very recently, the field has seen the leveraging of *Gaussian process regression* and *deep neural networks* into physics-informed machine learning [94, 95, 96, 97, 98, 99, 100]. For *Gaussian process regression*, the partial differential equation is encoded in an informative function prior; for *deep neural networks*, the partial differential equation induces a new neural network coupled to the standard uninformed data-driven neural network, see 3. We refer to this coupled data-partial differential equation deep neural network as a *physics-informed neural network*. New approaches, for example using generative adversarial networks, will be useful in the further development of *physics-informed neural networks*, for example, to solve stochastic partial differential equations, or fractional partial differential equations in systems with memory.

> ***Gaussian process regression*** *enables the creation of computationally inexpensive surrogates in a Bayesian approach. These surrogates do not assume a parametric form a priori and, instead, let the data speak for themselves. A significant advantage of Gaussian process surrogates is the ability to not only predict the response function in the parameter space, but also the associated epistemic uncertainty. Gaussian process regression has been used to creating surrogate models to characterize the effects of drugs on features of the electrocardiogram [106] and the effects of material properties on the stress profiles from reconstructive surgery [60].*

*Multiscale modeling for biological systems.* Biological materials are known to have a complex hierarchy of structure, mechanical properties, and biological behavior across spatial and temporal scales. Throughout the past two decades, modeling these multiscale phenomena has been a point of attention, which has advanced detailed deterministic models and their coupling across scales [26]. Strategies for multiscale modeling can be top down or bottom up, including: i) models of representative volume elements for the microscopic scale coupled with the larger spatial scales through boundary conditions in terms of the first derivatives of the macroscale fields, first order coupling, or higher derivatives, higher order coupling [36, 53, 111]; ii) micromechanics approaches [75, 93]; iii) reduced order and simplified models for upscaling [112, 123]; iv) scale-bridging or quasi-continuum methods [115, 133]. The explicit coupling of scales through representative volume elements is widely used because it can include many of the details and model complexities of the microscale, but it requires nested function evaluations such as nested finite element simulations or FE$^2$ that can easily become computationally intractable [21, 35, 52]. An additional complication in modeling biological systems comes from the inherent source of uncertainty in living matter, the high heterogeneity of the microscale, inter-subject variability, and stochastic nature of biological regulatory networks [16, 38, 68].

*Machine learning for multiscale systems.* Machine learning methods have recently permeated into composites research and materials design for example to enable the homogenization of representative volume elements with neural networks [94, 63, 65, 55] or the solution of high-dimensional partial differential equations with deep learning methods [41, 32, 33, 125, 126]. *Uncertainty quantification* in material properties is also gaining relevance, with examples of Bayesian model selection to calibrate of strain energy functions [71, 77] and uncertainty propagation with Gaussian processes of nonlinear mechanical systems [59, 60, 105]. These trends for non-biological systems point towards immediate opportunities for integrating machine learning and multiscale modeling in biological, biomedical, and behavioral systems and open new perspectives unique to the living nature of biological systems.

### 3.2. Applications and opportunities

*From localization to homogenization.* This application of machine learning in multiscale modeling is obvious, yet still unaccomplished. Leveraging data-driven approaches, the major objectives are localization and homogenization of information across the scales. The localization, the mapping of quantities from the macroscale, for example tissue stress or strain, to quantities at the microscale, for example cellular force or deformation, is crucial to understand the mechanobiology of the cell. The homogenization, the identification of the constitutive behavior at macroscale from the detailed behavior of representative units at the microscale, is critical to embed this knowledge into tissue or organ level simulations. In biological, biomedical, and behavioral systems, localization and homogenization present additional challenges and opportunities including the high-dimensional parameter space and inherent uncertainty, features that apply to both ordinary and partial differential equation based models, and the high degree of heterogeneity and microstructural complexity, features that mainly affect partial differential equation based models.

*From single source to multi-modality and multi-fidelity modeling.* Figure 4, left, illustrates various sources of data that that can potentially be combined with machine learning techniques. Biological, biomedical, and behavioral research crucially rely on experiments from different systems including in vitro culture, in vivo animal models and human data, and in silico experiments. The underlying assumption is that the data associated with each type of experiment or model are strongly correlated, even if they are not originating from the same system. Learning from one type of experiment or computer model can be used to improve the prediction at a higher level of fidelity for which information is scarce or difficult to obtain. Nonlinear multi-modality data fusion is a new way to combine information of various sources towards creating predictive models [88]. A typical example is integrating multi-omics data with biophysical models.

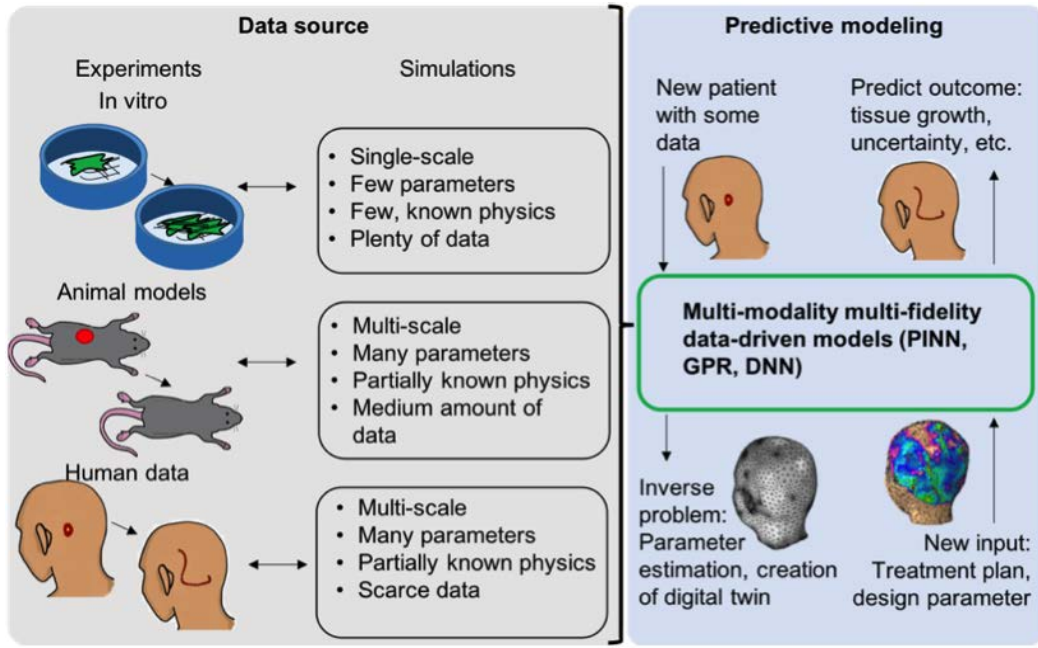

Figure 4: **Multi-modality and multi-fidelity modeling of biomedical systems.** Data from both experiments and computational models can be combined through machine learning to create predictive models. The underlying assumption is that, for a system of interest, data from different sources is correlated and can be fused. Parameter estimation, system identification, and function discovery result in inverse problems, for example, the creation of a digital twin, and forward problems, for example, treatment planning.

*From parameter estimation to system identification to function discovery.* Figure 4, right, illustrates the combination of parameter estimation, *system identification*, and function discovery required to create a digital twin. The combination of multi-modality, *multi-fidelity*, data-driven techniques allows us to create a personalized computational model for an individual by combining previous observations from multi-scale simulations, experiments, and clinical data with continuously updated recordings of this individual. Using the digital twin, we can probe different treatment scenarios and screen the design parameter space to create personalized treatment plans.

*From theoretical models to systems biology.* Living matter is characterized by its unique ability to respond and adapt to its environment. This can involve metabolic changes, inflammation, or mechanical changes such as growth and remodeling. Regulation of tissue activity is ultimately encoded in complex cell-signaling regulatory networks that operate at multiple spatial and temporal scales [118]. Modeling tissue adaptation thus involves accounting for classical equilibrium principles, e.g., momentum and energy, as well as signaling network dynamics often described by stochastic reactive transport models with many variables and parameters. The complexity of the system often defies intuition [51]. Machine learning could enable discovery of reactions, e.g., coagulation cascades, the solution of inverse problems for parameter estimation, and the quantification of biological uncertainty in model predictions. Figure 5 shows a possible application of predicting growth and remodeling at the tissue level based on cell-level information.

*From theoretical models to clinical applications.* Application of physics-based modeling in clinical practice is currently hindered by the difficulty to generate patient-specific predictions. Creating personalized models is time consuming and requires expert input and many different types of data, from multi-omics to biomechanical properties and measurements. On the opposite end, generic models are useful to understand mechanisms but are not suitable for planning individual interventions. Thus, there is an opportunity for machine learning—and transfer learning in particular—to generate individualized models or predictions in a fast and reliable manner without the need to create individualized models from scratch. Applications could include predicting the course of dissections in aortic dissections or quantifying wall shear stresses in aneurysms near the arterial wall [99]. Other applications include making predictions of whether thrombosis embolization will occur, or predictions of other cardiovascular diseases using multi-modality measurements integrated via *multi-fidelity* modeling to train neural networks instead of intricate models. Another possible application is optimization of surgery or prosthetic device design by combining off-line generic simulations with online data acquisition.

### 3.3. Open questions

*Modeling high-dimensional systems.* Can we model systems with high-dimensional input and hundreds of parameters? Biological systems are characterized by heterogenous microstructures, spatial heterogeneity, many constituents, intricate and noisy cell-signaling networks, and inherent variability across subjects. Attempts to model these systems necessarily rely on a high-dimensional parametric input space. Despite the recent progress in *uncertainty quantification* methods and smart sampling techniques using sparse grids and polynomial chaos methods, handling data in high-dimensional spaces remains challenging. However, deep learning techniques can exploit the compositional structure of approximating functions and can, in principle, beat the curse of dimensionality [89]. Generative adversarial networks can also be useful for effective modeling of parameterized partial differential equations with thousands of uncertain parameters [142, 143].

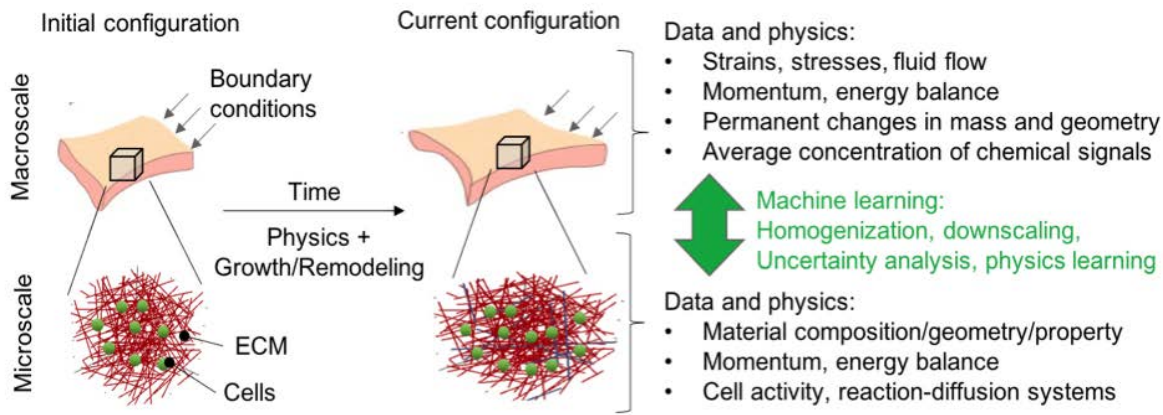

Figure 5: **Machine learning for multiscale modeling of biomedical systems.** Tissues are characterized by hierarchical structure across spatial and temporal scales associated with inherent variability. At both the macroscale and microscales, biological systems satisfy physics-based partial differential equations for mass, momentum, and energy balance. In addition, living systems have the unique ability to grow and remodel over time. This introduces an inherent coupling of the phenomena at the cellular and tissue scales. Machine learning enables the seamless integration of scales.

*Managing ill-posed problems.* Can we solve ill-posed inverse problems for parameter estimation or system identification? Many of the inverse problems for biological systems are ill posed, for example parameter estimation or *system identification*; they constitute boundary value problems with unknown boundary conditions. Classical mathematical approaches are not suitable in these cases. Methods for backward *uncertainty quantification* could potentially deal with the uncertainty involved in inverse problems, but these methods are difficult to scale to realistic settings. In view of the high dimensional input space and the inherent uncertainty of biological systems, posing inverse problems is challenging. For instance, it is difficult to determine if there are multiple solutions or no solutions at all, or to quantify the confidence in the prediction of an inverse problem with high-dimensional input data. The inherent regularization in the loss function of neural networks allows us to deal with ill-posed inverse partial differential equations without boundary or initial conditions and to discover hidden states and biophysics not possible with classical methods. Moreover, advances in probabilistic programming offer a promising path for performing scalable statistical inference for large-scale inverse problems with a large number of uncertain parameters.

*Discretizing space and time.* Can we remove or automate the tyranny of grid generation of conventional methods? Discretization of complex and moving three-dimensional domains remains challenging. It generally requires specific expertise and many hours of dedicated labor, and has to be re-done for each particular model. This is particularly important when creating personalized models with complex geometries and multiple spatial and temporal scales. While many efforts in machine learning are devoted to solving partial differential equations in a given domain, new opportunities include the use of machine learning to deal directly with the creation of the discrete problem. This includes automatic mesh generation, meshless interpolation, and parameterization of the domain itself as one of the inputs for the machine learning algorithms. Interestingly, some recent approaches with *physics-informed neural networks* entirely remove the notion of a mesh, and instead evaluate the conservation laws of mass, momentum, and energy at random points that are neither connected through a regular lattice nor through an unstructured grid.

> **Physics-informed neural networks** *are neural networks that solve supervised learning tasks while respecting physical constraints or encoding the partial differential equation in some way, for example, through the loss function. This technique is particularly powerful when dealing with sparse data from systems that obey known physical principles. Examples in biomedicine include diagnosing cardiovascular disorders non-invasively using four-dimensional magnetic resonance images of blood flow and arterial wall displacements [50], creating computationally efficient surrogates for velocity and pressure fields in intracranial aneurysms [99], and using nonlinear wave propagation dynamics in cardiac activation mapping [109].*

*Combining deterministic and stochastic models.* Can we couple conventional physics, mass and momentum balance, with stochastic reaction-diffusion over time to model the adaptation of living systems? While the laws that govern the physics of biological systems from the cellular scale to the tissue scale can be considered deterministic, the cell-signaling networks on the sub-cellular scales are inherently noisy [38, 119, 113]. We can model their multiphysics and multi-rate dynamics using neural networks by sharing the parameter space of coupled but separate neural networks, with each net representing a different multiscale process. This approach could alleviate some of the difficulties associated with stiff systems by exploiting the use of proper regularization terms in the loss function.

*3.4. Potential challenges and limitations*

***Integrating multi-modality training data.*** Even though nonlinear data fusion algorithms are currently being developed and improved, in the case of biological systems, as Figure 3 left suggests, the data can come from vastly different sources. A potential challenge is to appropriately weight the data from these different modalities to improve predictions. For instance, it remains a challenge to quantify to what extent an in vitro model or an animal model is representative of the human response. It is possible that even the most powerful machine learning tools cannot substantially improve predictions needed in the clinical setting without high quality data of the human-specific response. Designing new composite neural networks that exploit correlations between multi-modality and *multi-fidelity* data is a top priority in the near future.

> ***Multi-fidelity learning*** is a *supervised learning* approach used to fuse data from different sources to create a surrogate that can outperform predictions based on a single data source. Often times there is plenty of inexpensive, low fidelity data, for example from a simplified computational model or a simple experiment. At the same time, our confidence in the accuracy of this model or experiment is relatively low. In contrast, there is typically sparse expensive, high fidelity data from more complex computational models or experiments. Multi-fidelity learning exploits the correlations between the different fidelity levels to make predictions in regions of the input space for which no high fidelity data exist, but low fidelity measurements are easy to acquire. Recent examples include simulating the mixed convection flow past a cylinder [88], skin growth in tissue expansion [61], and cardiac electrophysiology [107].

***Tuning physics inspired neural networks for stiff multiscale systems.*** Multiscale systems lead to very complex landscapes of the loss function that current optimization solvers cannot deal with. Hence it could be impossible to train such systems. A possible approach is to use classical domain decomposition methods to deal with different regimes of scales and design corresponding distributed *physics-informed neural networks* that can be trained more easily and more accurately.

***Increasing rigor and reproducibility.*** The predictive power of models built with machine learning algorithms needs to be thoroughly tested. An important challenge is the creation of rigorous validation tests and guidelines for computational models. The use of open source codes and data sharing by the machine learning community is a positive step, but more benchmarks and guidelines are required for *physics-informed neural networks*. There is also an urgent need of sets of benchmark problems for biological systems. Reproducibility has to be quantified in terms of statistical metrics, as many optimization methods are stochastic in nature and may lead to different results.

***Knowing limitations.*** The main limitation of solving partial differential equations with neural networks and adversarial networks is the training time that results from solving non-convex optimization problems in high-dimensional spaces. This is a non-deterministic polynomial-time hard computational problem and one that will probably not be resolved in the near future. A possible solution is properly selecting the size of the system using domain decomposition techniques. Tuning the network parameters is a tedious and empirical job, but we can potentially adapt meta-learning methods that have been successfully used for *classification* problems to solve *regression* problems of partial differential equation modeling. Effective initialization, tailored to biological, biomedical, and behavioral systems, is a promising approach to alleviate some of these challenges and limitations.

> ***Classification*** is a *supervised learning* approach in which the algorithm learns from a training set of correctly classified observations and uses this learning to classify new observations, where the output variable is discrete. Examples in biomedicine include classifying whether a tumor is benign or malignant [34], classifying the effects of individual single nucleotide polymorphisms on depression [7], the effects of ion channel blockage on arrhythmogenic risk in drug development [108], and the effects of chemotherapeutic agents in personalized cancer medicine [27].

## 4. Data-driven approaches

Effectively an extension of statistics, machine learning is a method for identifying correlations in data. Machine learning techniques are function approximators and not predictors. This distinguishes them immediately from multiscale modeling techniques, which do provide predictions that can be based on parameter changes suggested by particular pathological or pharmacological changes [47]. The advantage of machine learning over both manual statistics and multiscale modeling is its ability to directly utilize massive amounts of data through the use of iterative parameter changes. This ability to handle big data is important in view of recent developments of ultra-high resolution measurement techniques like cryo-EM, high-resolution imaging flow cytometry, or four-dimensional-flow magnetic resonance imaging. Machine learning also allows us to analyze massive amounts of health data from wearable devices and smartphone apps [45]. As a data-mining tool, machine learning can help us bring experiment and multiscale modeling closer together. Machine learning also allows us to leverage data to build artificial intelligence applications to solve biomedical problems [128]. Figure 6 illustrates a framework for integrating multiscale modeling and machine learning in data-driven approaches.

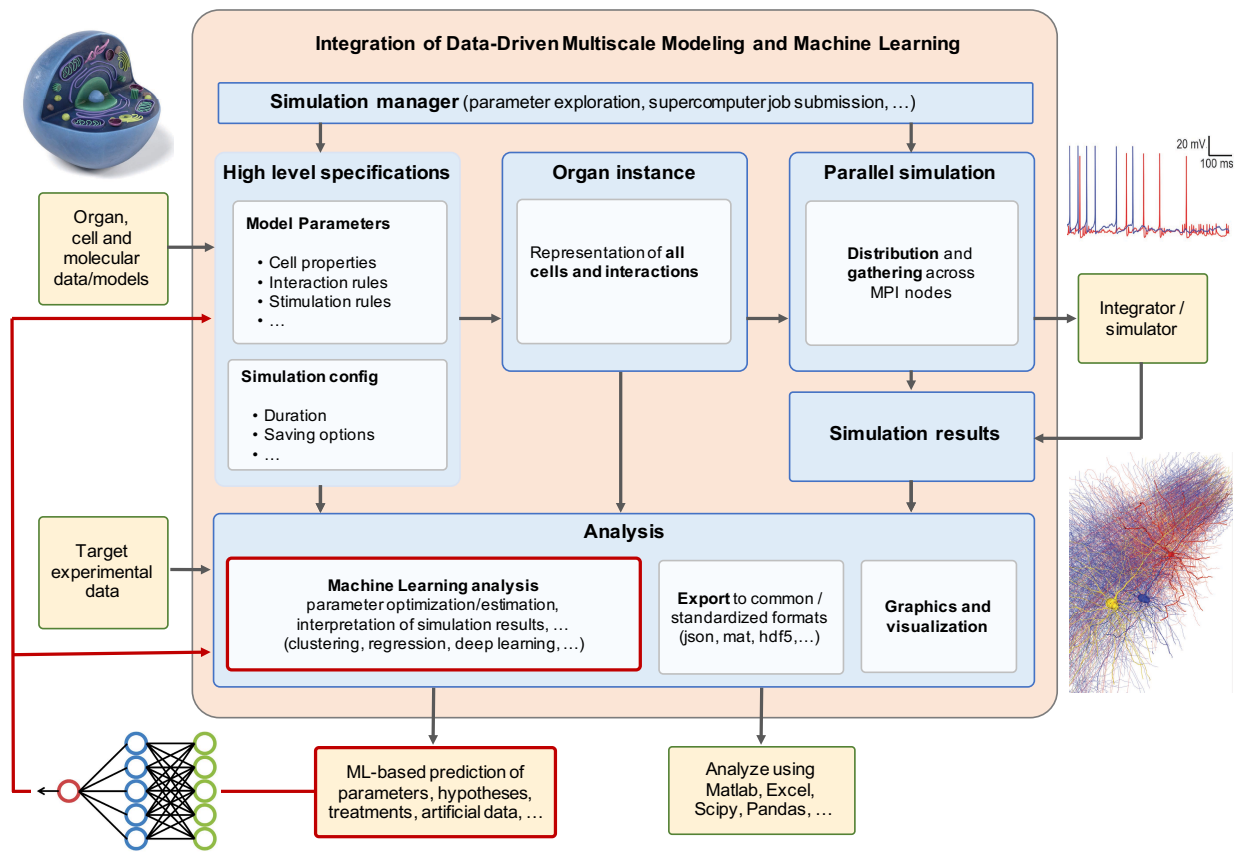

Figure 6: **Data-driven machine learning for multiscale modeling of biomedical systems.** Using clustering, regression, dimensionality reduction, reinforcement learning, deep learning, and parameter identification to analyze molecular, cellular, and organ level data.

*4.1. State of the art*

Most existing machine learning techniques identify correlations but are agnostic as to causality. In that sense, multiscale modeling complements machine learning: Where machine learning identifies a correlation, multiscale modeling can find causal mechanisms or a mechanistic chain [69]. Personalized medicine, where each patient's disease is considered a unique variant, can benefit from multiscale modeling to follow the particular parameters unique to that patient. Personalized models can then be based on individual differences measured by imaging [70, 90], by genomic or proteomic measures in the patient, or be based on the genomes of infectious agents or tumor cells. This will help in creating digital twins [72], models that incorporate both machine learning and multiscale modeling, for an organ system or a disease process in an individual patient. Using digital twins, we can identify promising therapies before trying them on the one patient [15]. As multiscale modeling attempts to leverage experimental data to gain understanding, machine learning provides a tool to preprocess these data, to automate the construction of models, and to analyze model output [132]. In the following, we focus primarily on applications of machine learning to multiscale modeling.

*4.2. Applications and opportunities*

Since machine learning involves computer-based techniques for predicting outcomes or *classification*s based on training data, all data-driven approaches, including mechanistic multiscale modeling, can benefit from application of machine learning.

***From simulation experiments to animal experiments.*** In biomedicine, the original area of big-data research was the identification of the human genome, which employed machine learning to construct consistent frames out of genome fragments. Since then, extensions of genomic studies involve large numbers of patients and controls in genome-wide association studies, which compare single nucleotide polymorphisms in patients versus controls [7, 56]. Single nucleotide polymorphisms adjacent to coding sequences suggest which gene product *might* be involved in the disease. Once particular proteins are identified, multiscale modeling can track the effects up through the scales from molecular to cellular to intercellular to organ and organism. In addition to describing dynamics across scales, multiscale modeling can identify how multiple gene products may interact to produce the disease, given that most diseases are polygenic rather than being caused by a single mutation. Simulation experiments can also identify how other allele combinations would produce different disease manifestations and identify degrees of penetrance of a mutation. Simulation can then help us plan animal experiments to add further evidence.

*From smaller scales to larger scales.* A major problem for multiscale modeling is the identification of appropriate model parameters. Ideally, multiscale modeling parameters are all based on consistent experimental measurements. Realistically, biological parameters may be measured in various species, in various cell types, at various temperatures, at various ages, and in different *in vivo* and *in vitro* preparations. Medical multiscale modeling, and medical conclusions, are applied to humans, but are often based on animal models. Additionally, many enzymes have large numbers of isoforms and different phosphorylation states that make generalization problematic. For all these reasons, it is typically necessary to identify the parameters on a multiscale model to ensure realistic dynamics at the higher scales of organization. Machine learning techniques have been used extensively to tune the parameters to replicate these higher-level dynamics. An example of this is the use of genetic algorithms and *evolutionary algorithms* in neural models [17, 30, 82]. Going beyond inference of parameters, recurrent neural networks have been used to identify unknown constitutive relations in ordinary differential equation systems [40].

> *Evolutionary algorithms* *are generic population-based optimization algorithms that adopt mechanisms inspired by biological evolution to generate new sampling points for further function evaluation. Strategies include reproduction, mutation, recombination, and selection. Evolutionary algorithms have been used successfully for automatic parameter tuning in multiscale brain modeling [30].*

*From multiscale modeling to machine learning.* Multiscale models can provide insight into a biological, biomedical, or behavioral system at a high level of resolution and precision. They can systematically probe different conditions and treatments, faster, more cost-effective, and often beyond what is possible experimentally. This parameter screening naturally produces massive output datasets, which are ideally suited to machine learning analysis. To no surprise, machine learning methods are progressively becoming part of the tool suite to analyze the output of multiscale models [31, 144]. A recent example is the study of *clustering* to study the effects of potential simulated treatments [64, 82].

> *Clustering* *is an unsupervised learning method that organizes members of a dataset into groups that share common properties. Typical examples in biomedicine include clustering the effects of simulated treatments [64, 82].*

### 4.3. Open questions

*Classifying simulation and experiment.* To what extent does the multiscale modeling differ from experiment? Machine learning tools allow us to *cluster* and *classify* the predictions of the multiscale model and systematically compare simulated and experimental datasets. Where simulation and experiment differ, machine learning can identify potential high-order features and suggest iterative refinements to improve the multiscale model.

*Identifying missing information.* Do the chosen parameters provide a basis set that allows production of the needed higher-scale model dynamics? Multiscale simulations and generative networks can be set up to work in parallel, alongside the experiment, to provide an independent confirmation of parameter sensitivity. For example, the circadian rhythm generators provide relatively simple dynamics but have very complex dependence on numerous underlying parameters, which multiscale modeling can reveal. We could then use *generative models* to identify both the underlying low dimensionality of the dynamics and the high dimensionality associated with parameter variation. Inadequate multiscale models could be identified with failure of generative model predictions.

> *Generative models* *are statistical models that capture the joint distribution between a set of observed or latent random variables. A recent study used deep generative models for chemical space exploration and matter engineering [110].*

*Creating surrogates.* How can we use generative adversarial networks to create new data sets for testing multiscale models? Conversely, how can we create training or test instances using multiscale modeling to use with deep learning models? A deep learning network could be deployed more widely and provide answers more quickly than a multiscale modeling dynamic simulation, permitting, for example, prediction of pharmaceutical efficacy for patients with particular genetic inheritance in personalized medicine. This would be particularly important for on-body digital twins which need to function rapidly with limited computational resources

*Identifying relevant processes and interactions.* How can we use machine learning to bridge scales? For example, machine learning could be used to explore responses of both immune and tumor cells in cancer based on single-cell data. A multiscale model could then be built on the families of solutions to codify the evolution of the tumor at organ- or metastasis-scale.

**Supplementing training data.** *Supervised learning*, as used in deep networks, is a powerful technique but requires large amounts of training data. Recent studies have shown that, in the area of object detection in image analysis, simulation augmented by *domain randomization* can be used successfully as a supplement to existing training data [131]. In areas where multiscale models are well-developed, simulation across many parameter has been used as a supplement to existing training data for nonlinear diffusion models to provide physics-informed machine learning [100, 121, 122]. Similarly, multiscale models can be used in biological, biomedical, and behavioral systems to augment insufficient experimental or clinical data sets. Machine learning can provide tools to verify the validity of the simulation results. Multiscale models can then expand the datasets towards developing machine learning and artificial intelligence applications.

> **Domain randomization** *is a technique for randomizing the field of an image so that the true image is also recognized as a realization of this space. Domain randomization has been used successfully to supplement training data [131].*

*4.4. Potential challenges and limitations*

**Developing new architectures and algorithms inspired by biological learning.** Chess and go are two games, difficult for humans, which have been solved successfully by artificial intelligence using deep learning. Deep learning has also shown success in image recognition, a function that utilizes large amounts of brain real estate [129]. By contrast, activities that real brains networks are very good at remain elusive. For example, the control systems of a mosquito engaged in evasion and targeting are remarkable considering the small neuronal network involved. This limitation provides opportunities for more detailed brain models to assist in developing new architectures and new learning algorithms. Incorporating spiking [49] or oscillatory dynamics at the mesoscopic or macroscopic levels could inspire novel low-energy architectures and algorithms. Deep learning and *reinforcement learning* were both motivated by brain mechanisms. Understanding biological learning has the potential to inspire novel and improved machine learning architectures and algorithms [44].

> **Reinforcement learning** *is a technique that circumvents the notions of supervised learning and unsupervised learning by exploring and combining decisions and actions in dynamic environments to maximize some notion of cumulative reward. Of broad relevance is understanding common learning modes in biological, cognitive, and artificial systems through the lens of reinforcement learning [13, 80].*

**Identifying disease progression biomarkers and mechanisms.** There are abundant challenges for data-driven approaches for integrating machine learning and multiscale modeling towards understanding and diagnosing specific disease states. If machine learning could identify predictive disease progression biomarkers, multiscale modeling could follow up to identify mechanisms at each stage of the disease with the ultimate goal to propose interventions that delay, prevent, or revert disease progression.

## 5. Theory-driven approaches

Theory-driven approaches aspire to answer the following questions: How can we leverage structured physical laws and mechanistic models as informative prior information in a machine learning pipeline towards advancing modeling capabilities and expediting the simulation of multiscale systems? Given imperfect and irregularly sampled data, how do we identify the form and parameters of a governing law, for example an ordinary or partial differential equation, and use it for forecasting? Figure 7 illustrates a closed-loop integration of theory-driven machine learning and multiscale modeling to accelerate model- and data-driven discovery.

*5.1. State of the art*

Theory-driven machine learning is both a decently mature and quickly evolving area of research. It is mature in that methods for learning parameters for a model such as dynamic programming and variational methods have been known and applied for a long time. Although these methods are generally not considered to be tools of machine learning, the difference between them and current machine learning techniques may be as simple as the difference between a deterministic and a stochastic search [140]. Dynamic programing and variational methods are very powerful when we know the form of the model and need to constrain the parameters within a specified range to reproduce experimental observations. Machine learning methods, however, can be very powerful when the model is completely unknown or when there is uncertainty about its form. Mixed cases can exist as well. For example, when modeling the dynamics of a cell, we may know the rate laws that need to be solved and obtain the rate parameters using an optimization algorithm; yet, which reactions are regulated under what conditions may be a mystery for which there are no adequate models.

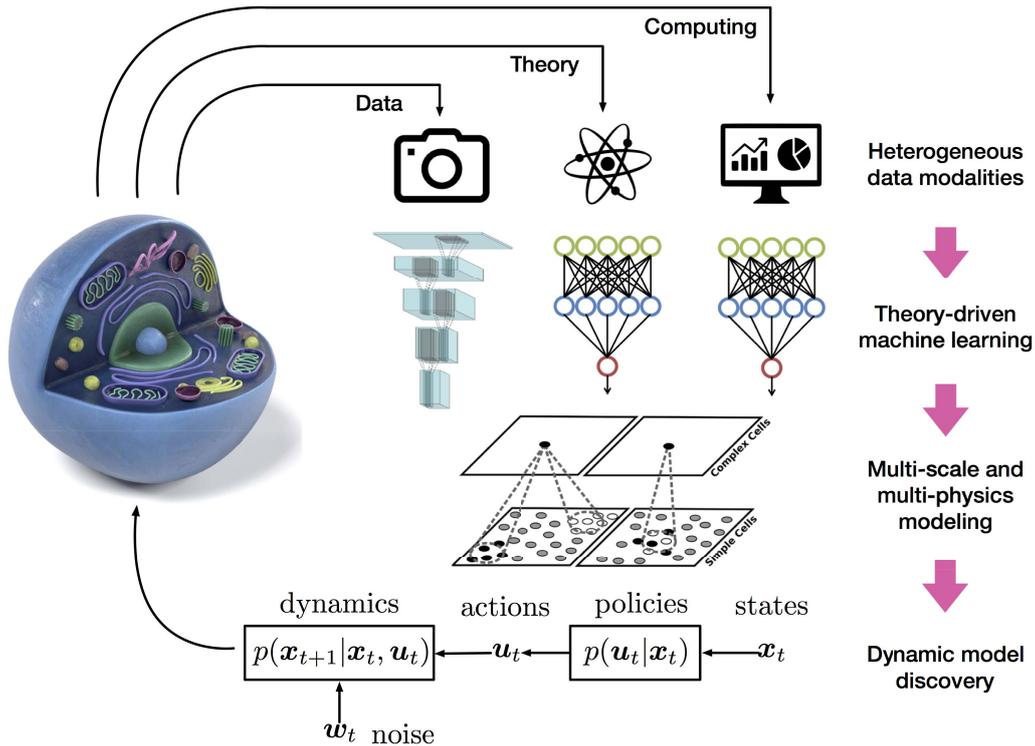

Figure 7: **Theory-driven machine learning for multiscale modeling of biological systems.** Theory-driven machine learning can yield data-efficient work-flows for predictive modeling by synthesizing prior knowledge and multi-modal data at different scales. Probabilistic formulations can also enable the quantification of predictive uncertainty and guide the judicious acquisition of new data in a dynamic model-refinement setting.

Theory-driven machine learning can enable the seamless synthesis of physics-based models at multiple temporal and spatial scales. For example, *multi-fidelity* techniques can combine coarse measurements and reduced order models to significantly accelerate the prediction of expensive experiments and large-scale computations [88]. In drug development, for example, we can leverage theory-driven machine learning techniques to integrate information across ten orders of magnitude in space and time towards developing interpretable classifiers that enable us to characterize the potency of pro-arrhythmic drugs [106]. Based on *Gaussian process regression*, these approaches can effectively explore the interplay between drug concentration and drug toxicity by probing the effect of different drugs on ion-channel blockage, cellular action potentials, and electrocardiograms using coarse and low-cost models, anchored by a few, judiciously selected, high-resolution simulations of cardiac electrophysiology [107].

Leveraging probabilistic formulations, theory-driven machine learning techniques can also inform the judicious acquisition of new data and actively expedite tasks involving the exploration of large parameter spaces or the calibration of complex models. For example, we could devise an effective data acquisition policy for choosing the most informative meso-scale simulations that need to be performed to recover detailed constitutive laws as appropriate closures for macroscopic models of complex fluids [145]. Building on recent advances in *automatic differentiation* [10], techniques such as *neural differential equations* [22] are also expanding our capabilities in calibrating complex dynamic models using noisy and irregularly sampled data.

> ***Automatic differentiation*** *is a family of techniques to efficiently and accurately evaluate derivatives of numeric functions expressed as computer programs. Not to be confused with symbolic or numerical differentiation, automatic differentiation is an exact procedure for differentiating computer code by redefining the semantics of the operators, composing a complex program to propagate derivatives per the chain rule of differential calculus [10]. Along with the use of graphics processing units, automatic differentiation has been one of the key backbones of the current machine learning revolution. It significantly expedits the prototyping and deployment of predictive data-driven models.*

> ***Neural differential equations*** *are machine learning models that aim to identify latent dynamic processes from noisy and irregularly sampled time-series data [22]. Leveraging recent advances in* automatic differentiation *[10], they can efficiently back-propagate through ordinary or partial differential equation solvers to calibrate complex dynamic models and perform forecasting with quantified uncertainty. Examples in biomedicine include predicting in-hospital mortality from irregularly sampled time-series containing measurements from the first 48 hours of a different patient's admission to the intensive care unit [103].*

More recently, efforts have been made to directly bake theory into machine learning practice [99]. This enables the construction of predictive models that adhere to the underlying physical principles, including conservation, symmetry, or invariance, while remaining robust even when the observed data are very limited. For example, a recent model only utilized conservation laws of reaction to model the metabolism of a cell. While the exact functional forms of the rate laws was unknown, the equations were solved using machine learning [24]. An intriguing implication of such theory-driven machine learning approaches is related to their ability to leverage auxiliary observations to infer quantities of interest that are difficult to measure in practice [100]. An example includes the use of *physics-informed neural networks* to infer the arterial blood pressure directly and non-invasively from four-dimensional magnetic resonance images of blood velocities and arterial wall displacements by leveraging the known dynamic correlations induced by first principles in fluid and solid mechanics [50]. A commom thread between these approaches is that they rely on conventional neural network architectures, including fully connected or convolutional, and constrain them using physical laws as penalty terms in the loss function that drives the learning process. An alternative, yet more laborious route is to design new neural architectures that implicitly remain invariant with respect to the symmetry groups that characterize the dynamics of a given system. A representative example of this school of thought can be found in covariant molecular neural networks; a rotationally covariant neural network architecture for learning the behavior and properties of complex many-body physical systems [6].

*5.2. Applications and opportunities*

***From models and patient data to personalized medicine.*** Theory-driven and computational methods have long aspired to provide predictive tools for patient monitoring, diagnostics, and surgical planning. However, high-fidelity predictive models typically incur a large computational cost and rely on tedious calibration procedures that render them impractical for clinical use. Theory-driven machine learning, for example in the form of *multi-fidelity* or *physics-informed* approaches, has the potential to bridge the gap between modeling, predictions, and clinical decision making by enabling the seamless and cost-effective integration of computational models and disparate data modalities, for example from medical imaging, laboratory tests, and patient records. In the age of the Digital Twin, this integration enables new capabilities for assimilating data from medical devices into predictive models that enable the assessment of health risks and inform preventative care and therapeutic strategies on a personalized basis.

***From protein biology to physics of the cell.*** In contrast to the clinical case, theory-driven computational methods have a long history of providing high-fidelity, predictive models for many fields in the basic sciences with protein biology as one of the most prominent examples [76, 135]. However, once we move to the scale where non-equilibrium phenomena occur, the rate parameters necessary to solve either the full mass action rate law or even approximations such as the Michealis-Menten equation have been difficult and expensive to obtain, as they depend on careful in vitro kinetic studies. The Costello-Garcia-Martin study mentioned above is a significant step, but the functional form of the rate laws, which describe the underlying physics remain unknown. Accordingly, a grand challenge application that is ripe for further development is combining theory-driven machine learning with multi-scale modeling to understand the physics of the cell, especially with a view towards the emergence of function. We could, for example, generate high-fidelity data from simulations based on maximum entropy assumptions, and then use these data to learn feasible solution spaces.

***From interpolation to extrapolation.*** When data can be generated from theory-driven models, machine learning techniques have enjoyed immense success. Such tasks are usually based on interpolation in that the input domain is well specified, and we have sufficient data to construct models that can interpolate between the dots. This is the regime where discriminative, black box methods such as deep learning perform best. When extrapolation is needed instead, the introduction of prior knowledge and appropriate inductive biases through theory-driven methods can effectively steer a machine learning algorithm towards physically consistent solutions. Theory-driven machine learning approaches present a unique opportunity for leveraging the domain knowledge and mechanistic insight brought by the multi-scale modeling community to develop novel learning algorithms with enhanced robustness, data-efficiency, and generalization performance in data-limited regimes. We anticipate such developments to be crucial for leveraging the full potential of machine learning in advancing multi-scale modeling for biological, biomedical, and behavioral systems.

*5.3. Open questions*

***Elucidating mechanisms.*** Can theory-driven machine learning approaches enable the discovery of interpretable models that can not only explain data, but also elucidate mechanisms, distill causality, and help us probe interventions and counterfactuals in complex multi-scale systems? For instance, causal inference generally uses various statistical measures such as partial correlation to infer causal influence. If instead, the appropriate statistical measure were known from the physics such as a statistical odds ratio from thermodynamics, would the causal inference be more accurate or interpretable as a mechanism?

***Understanding the emergence of function.*** Understanding the emergence of function is of critical importance in biology and medicine, environmental studies, biotechnology and other biological sciences. The study of emergence necessitates the ability to model collective action on a lower scale to predict how the phenomena on the higher scale emerge from the collective action. Can theory-driven machine learning, combined with sparse and indirect measurements of the phenomena, produce a mechanistic understanding of how biological phenomena emerge?

*Exploring massive design spaces.* Can theory-driven machine learning approaches uncover meaningful and compact representations for complex inter-connected processes, and, subsequently, enable the cost-effective exploration of vast combinatorial spaces? A typical example is the design of bio-molecules with target properties in drug development.

*Predicting uncertainty.* *Uncertainty quantification* is the back-bone of decision making. Can theory-driven machine learning approaches enable the reliable characterization of predictive uncertainty and pinpoint its sources? The quantification of uncertainty has many practical applications such as decision making in the clinic, the robust design of synthetic biology pathways, drug target identification and drug risk assessment, as well as guiding the informed, targeted acquisition of new data.

*Selecting the appropriate tools.* Is deep learning necessary in theory-driven learning? In principle, the more domain knowledge is incorporated into the model the less needs to be learned and the easier the computing task will become. More knowledge will enable researchers to take on even greater challenges, which, in turn, may require more learning. It is likely that the applications will utilize a range of techniques, from dynamic programming to variational methods to standard machine learning to deep learning. Is high performance computing required when theory-driven models are employed? The answer here probably depends on the application and the depth of the model used in learning, with the larger the multi-scale model, the more computing necessary.

*5.4. Potential challenges and limitations*

A major challenge in theory-driven approaches towards understanding biological, biomedical, and behavioral systems, is obtaining sufficient data to answer the driving question of interest.

*Combining low- and high-resolution data.* Can theory-driven machine learning be utilized to bridge the gap between qualitative 'omics data and the quantitative data needed for prediction? For example, RNASeq data has become fairly quantitative, but the amplification of transcripts using polymerase chain reaction can add uncertainty to the final measures. Proteomics and metabolomics assays can be quite quantitative when using nuclear magnetic resonance or multiple reaction monitoring [54], but nuclear magnetic resonance has a relatively narrow dynamic range for quantification and multiple reaction monitoring mass spectrometry is not high throughput. Similarly, isotope labeling studies such as metabolic flux analysis [139] and absolute quantitation by mass spectroscopy [11, 67, 85] provide highly valuable information, but are low-throughput and relatively costly. High throughput methods such as shotgun proteomics [141] or global metabolomics determine whether a protein, peptide, or metabolite is observed, but not whether it is actually present since different species have different detectability characteristics. For these reasons, the use of high-throughput biological data in machine learning remains a challenge, but combining theory-driven approaches with *multi-fidelity* data would help reduce the uncertainty in the analysis.

*Minimizing data bias.* Can arrhythmia patients trust a neural net controller embedded in a pacemaker that was trained under different environmental conditions than the ones during their own use? Training data come at various scales and different levels of fidelity. These data are typically generated by existing models, experimental assays, historical data, and other surveys, all of which come with their own biases. As machine learning algorithms can only be as good as the data they have seen, proper care needs to be taken to safe-guard against biased models and biased data-sets. Theory-driven approaches could provide a rigorous foundation to estimate the range of validity, quantify the uncertainty, and characterize the level of confidence of machine learning based approaches.

*Knowing the risk of non-physical predictions.* Can new data fill the gap when the multi-scale model lacks a clean separation between the fast and slow temporal scales or between the small and large spatial scales? From a conceptual point of view, this is a problem of supplementing the set of physics-based equations with constitutive equations, an approach, which has long been used in traditional engineering disciplines. While data-driven methods can provide solutions that are not constrained by preconceived notions or models, their predictions should not violate the fundamental laws of physics. Sometimes it is difficult to determine whether the model predictions obey these fundamental laws. This is especially the case when the functional form of the model cannot be determined explicitly, for instance in deep learning. This makes it difficult to know whether the analysis predicts the correct answer for the right reasons. There are well-known examples of deep learning neural networks that appear to be highly accurate, but make highly inaccurate predictions when faced with data outside their training regime [83], and others that make highly inaccurate predictions based on seemingly minor changes to the target data [120]. Integrating machine learning and multiscale models that a priori satisfy the fundamental laws of physics would help address this limitation.

## 6. Conclusion

Many exciting new applications emerge at the interface of machine learning and multiscale modeling. Immediate applications in multiscale modeling include system identification, parameter identification, sensitivity analysis, and uncertainty quantification and applications in machine learning are physics-informed neural networks. Integrating machine learning and multiscale modeling can have a massive impact in the biological, biomedical, and behavioral sciences, both in diagnostics and prognostics, and this review has only touched the surface. Undeniably, applications become more and more sophisticated and have to be increasingly aware of the inherent limitations of overfitting and data bias. A major challenge to make progress in this field will be to increase transparency, rigor, and reproducibility. We hope that this review will stimulate discussion within the community of computational mechanics and reach out to other disciplines including mathematics, statistics, computer science, artificial intelligence, and precision medicine to join forces towards personalized predictive modeling in biomedicine.

## Acknowledgements

The authors acknowledge stimulating discussions with Grace C.Y. Peng, Director of Mathematical Modeling, Simulation and Analysis at NIBIB, and the support of the National Institutes of Health grants U01 HL116330 (Alber), R01 AR074525 (Buganza Tepole), U01 EB022546 (Cannon), R01 CA197491 (De), U24 EB028998 (Dura-Bernal), U01 HL116323 and U01 HL142518 (Karniadakis), U01 EB017695 (Lytton), R01 EB014877 (Petzold) and U01 HL119578 (Kuhl), as well as DARPA grant HR0011199002 and Toyota Research Institute grant 849910, (both Garikipati). This work was inspired by the 2019 Symposium on Integrating Machine Learning with Multiscale Modeling for Biological, Biomedical, and Behavioral Systems (ML-MSM) as part of the Interagency Modeling and Analysis Group (IMAG), and is endorsed by the Multiscale Modeling (MSM) Consortium, by the U.S. Association for Computational Mechanics (USACM) Technical Trust Area Biological Systems, and by the U.S. National Committee on Biomechanics (USNCB). The authors acknowledge the active discussions within these communities.